\documentclass[preprint,authoryear,12pt]{elsarticle}
\usepackage{graphicx}
\usepackage{a4wide}
\usepackage{epstopdf}
\usepackage{booktabs}
\usepackage{natbib} 
\usepackage{multirow}
\usepackage{amssymb}
 \usepackage{amsthm}
 \usepackage{amsmath}
  \usepackage{lscape}
\usepackage{hyperref}

\journal{Energy Economics}

\begin{document}

\begin{frontmatter}

\title{Rockets and feathers meet Joseph: Reinvestigating the oil-gasoline asymmetry on the international markets}

\author[utia,ies]{Ladislav Kristoufek} \ead{kristouf@utia.cas.cz}
\author[ies]{Petra Lunackova}

\address[utia]{Institute of Information Theory and Automation, Academy of Sciences of the Czech Republic, Pod Vodarenskou Vezi 4, 182 08, Prague, Czech Republic, EU} 
\address[ies]{Institute of Economic Studies, Faculty of Social Sciences, Charles University in Prague, Opletalova 26, 110 00, Prague, Czech Republic, EU}

\begin{abstract}
We reinvestigate the ``rockets and feathers'' effect between retail gasoline and crude oil prices in a new framework of fractional integration, long-term memory and borderline (non-)stationarity. The most frequently used error-correction model is examined in detail and we find that the prices return to their equilibrium value much more slowly than would be typical for the error-correction model. Such dynamics is usually referred to as ``the Joseph effect''. The standard procedure is shown to be troublesome and we introduce three new tests to investigate possible asymmetry in the price adjustment to equilibrium under these complicated time series characteristics. On the dataset of seven national gasoline prices, we report that apart from Belgium, there is no asymmetry found. The proposed methodology is not limited to the gasoline and crude oil case but it can be utilized for any asymmetric adjustment to equilibrium analysis. 
\end{abstract}

\begin{keyword}
rockets and feather \sep asymmetry \sep gasoline \sep crude oil \sep cointegration \\
\textit{JEL codes:} Q40, Q43, Q48
\end{keyword}

\end{frontmatter}

\newpage

\section{Introduction}

Gasoline prices shoot up like rockets and fall down slowly like feathers -- such is a popular belief and feeling of retail customers at gasoline stations. Increasing gasoline prices in the last decade have made such notion even more relevant to general public as well as to policy makers. The study of \cite{bacon1991rockets} has coined the term of ``rockets and feathers''  into the literature and since then, the topic has attracted much attention. The price of gasoline, after controlling for taxes, is primarily driven by the crude oil prices, even though such effect is indirect as there are usually several steps from the oil rigs and wells to the retail customers. Although the passthrough of the oil price to the retail gasoline prices might take relatively a long time due to economic reasons such as transportation, menu costs, storage and others, the price adjustment should be symmetric whether the oil prices are going up or down. \cite{Mandelbrot1968} refer to such long-term dynamics as the Joseph effect illustrative by the biblical story of Joseph son of Jacob who interpreted a dream of the Egyptian pharaoh about upcoming seven years of plenty followed by seven years of famine (Chapter 41 of the Book of Genesis). The dream-telling had been rewarded and Joseph served as the pharaoh's vizier. The years of plenty and the years of famine represent long periods when time series are above or below their long-term mean. From the econometric standpoint, this is represented by a slow decay of autocorrelation function of the series characteristic for the long-term correlated (long-range correlated, or persistent) series \citep{Beran1994,Samorodnitsky2006}.

Even though the parallel between price adjustment and the Joseph effect might be vivid and straightforward, it does not reflect the approach taken in majority of the empirical literature investigating the ``rockets and feathers'' effect in the gasoline market. In Section 2, we present a comprehensive literature review of the asymmetric price adjustment between gasoline and crude oil and we show that the studies usually begin with the assumption of the long-term equilibrium relationship between retail gasoline (or diesel in some cases) and crude oil. Specifically, the cointegration relationship is being built upon. This is well grounded both theoretically and empirically. However, the next step usually stems in estimating some form of an error-correction model. The deviation from equilibrium, represented by the error-correction term in the cointegration equation, is thus assumed to return to zero, i.e. the equilibrium state, rather quickly. We describe the cointegration and error-correction model methodology in Section 3. There, we also introduce the analyzed dataset, which comprises of the gasoline markets of Belgium, France, Germany, Italy, the Netherlands, the UK and the USA, and we focus on the basic dynamic properties of the series as well. We show that the gasoline markets are indeed cointegrated with the crude oil. However, we also show that the gasoline prices return to their long-run equilibrium very slowly. Specifically, we show that such dynamics can be attributed to the long-term memory and hence the Joseph effect rather than the rapidly adjusting error-correction model. We argue that such strong memory makes the standard error-correction models and their variants infeasible and as a solution, we propose three new tests for examining asymmetry in the cointegration framework. In Section 4, we present results of the asymmetry testing in the international gasoline markets and we show that apart from Belgium, there is no statistical evidence of the ``rockets and feathers'' dynamics towards equilibrium, and we also outline possible directions of future research in this area. Section 5 concludes.

\section{Literature review} 

The term ``rockets and feathers'' has been connected with crude oil and retail gasoline since 1991 when Robert Bacon published his famous article \citep{bacon1991rockets}. Since then, vast research focusing on the (a)symmetric behavior of prices ``at the pump'', has been performed. Its motivation is to explain this phenomenon and understand whether any policy would improve the current market situation. As the literature on the topic is quite broad, we summarize the reviewed articles in Tab. \ref{tabLR} while focusing mainly on the analyzed time period, location and possible asymmetry.

The most common econometric approach investigating the asymmetry is the error-correction model (ECM) and we focus on this dominant branch of the literature. All the ECMs are based on the two step \cite{Engle1987} procedure that exploits the long-run equilibrium relationship between, in our case mostly, crude oil and retail gasoline. Various ECM specifications could be put into three groups -- asymmetric ECM (used by most studies), threshold autoregressive ECM \citep{godby2000testing,al2007retail} and ECM with threshold cointegration \citep{chen2005threshold}. For more detailed analysis, see the work of \cite{grasso2007asymmetric} who research the sensitivity of various ECM models in order to understand how the choice of a particular model influences the results.

Existing literature differs by country, sample period and data frequency, econometric model and research question. Paper of \cite{borenstein1997gasoline} has influenced all subsequent papers and it serves as the reference point until now. The study is focused on the US market in 1986-1992 and its findings are based on ECM. The authors provide evidence for a common belief that after a crude oil price changes, gasoline prices rise faster than they fall. They try to identify the stage where the asymmetry occurs but is seems to be spread over all stages. The paper also offers an explanation for the asymmetric retail price adjustment (sticky prices, production lags, and inventories). 

\cite{balke1998crude} extend the previous study using several different model specifications and they confirm the asymmetry and conclude that the findings are sensitive to model specifications but not to the sample period. 
\cite{bachmeier2003new} use daily (spot) prices from the US market and find no evidence of asymmetry in wholesale gasoline prices. Analysis of \cite{borenstein1997gasoline} is performed on weekly and biweekly data and that is how \cite{bachmeier2003new} explain different results -- broader interval can result in a significant bias. 

The literature on the ``rockets and feathers'' phenomenon can be viewed and compared from many different angles. 

Firstly, the studies can be separated according to a country of interest. Most of the studies focus on the US market, some on Canada and the UK, few on Western European countries, other countries like Chile \citep{balmaceda2008asymmetric} or New Zealand \citep{liu2010there} are rare. According to \cite{duffy1996retail}, the asymmetric effect depends also on the market size, and conclusions made based on local markets' data cannot be generalized and applied to national markets. \cite{deltas2008retail} also relates the asymmetry to the local market conditions. Secondly, according to the objective, the articles' aim is to (dis)prove the asymmetry or to analyze the asymmetry itself. Thirdly, sample period and data frequency matter, and mainly the latter one that varies from daily to monthly, and various specifications (simple price averages or prices collected on the specific day of the week) are utilized. For example, \cite{bettendorf2003price} estimated the ECM for five datasets, one for each working day, to find out whether the choice of the day of week matters. Fourthly, according to the results, asymmetry is prevailing but it is not unanimous. \cite{godby2000testing} work on Canadian data and, together with \cite{bachmeier2003new} and \cite{karrenbrock1991behavior}, they are among few authors who cannot reject symmetry. The three mentioned studies that found no asymmetry all worked with different data frequency which suggests that frequency may not be the crucial factor. Some findings are also neutral as in \cite{bettendorf2003price} or \cite{oladunjoye2008market}.
From a different angle, \cite{douglas2010gasoline} claims that the asymmetry found is caused by the outliers in data. The asymmetry disappears after exclusion of the outliers.

Last but not least, we can split the articles according to the approach that explains the asymmetry as all papers discuss the causes of the asymmetry as well. There are three major explanations. 

The first one focuses on market power and connects the phenomenon to oligopolistic theory. Market power is the most widespread explanation. Price of retail gasoline is easily available and of interest to all drivers, which is a large group of consumers that frequently suspect some form of collusion, even though there is little evidence of market power abuse \citep{brown2000gasoline}. Moreover, even if there was a player with significant market power, \cite{peltzman2000prices} does not find a link between market power and asymmetric pricing.
On the contrary, \cite{radchenko2005oil} attributes asymmetry to the oligopolistic theory and finds negative relation between oil price volatility and asymmetric response of gasoline prices -- the degree of asymmetry declines with increase in oil price volatility. 

The second explanation analyzes the demand side and claims that consumers cause part of the asymmetry, theory known as consumer search theory. Consumers search less intensively for a better deal when prices are falling. Imagine a driver passing by a gas station who spots the gasoline rack prices and now gasoline costs less than he expected. If our hypothetical driver is in need of gasoline, he will stop at that station (and observe others' prices as he goes his way). In his theoretical paper, \cite{tappata2009rockets} suggests that asymmetric response emerges naturally, based on consumer search. \cite{lewis2011asymmetric} also says that consumers search less when prices are falling, the reduced search causes a slower price response. 
\cite{johnson2002search} gives the following implication -- if search costs are such an important factor that determines the lag length, then there should be a shorter adjustment lag in case of diesel than in case of gasoline, as diesel is typically bought in larger quantities and more frequently, therefore the customers have a greater incentive to search.

Other (minor) explanations form the third group. Decreasing inventories are the reason to either produce less (resulting in price increase) or buy more inputs (resulting in price increase as well). Unfortunately, opposite does not have to hold for increasing amount of inventories which adjust more slowly. The intention is to avoid abrupt price changes and not to increase already high price volatility. The FIFO (first in first out) accounting principle built in the pricing process does not smooth the price/costs changes and refinery adjustment costs either, it follows the behavior of inventories. Gasoline prices respond to cost shocks with lag in order to spread the adjustment costs \citep{borenstein2002sticky}.

\cite{panagiotidis2007oil} test the assumption that liberalization should cause decoupling of gas and oil prices on the British data. Their results do not support the expectations, which on the contrary supports the cointegration relation of crude oil and retail gasoline. \cite{galeotti2002rockets} revisit the phenomenon analyzing an international data set (joint data for France, Spain, Italy, Germany and the United Kingdom) and break up the process into two stages -- refinery and distribution. In both cases, asymmetry is found. \cite{verlinda2008rockets} then focuses on local market, believing in its greater information value. Employing detailed weekly data at a station level and local market characteristics, the author concludes that the degree of asymmetry is influenced by brand identity, proximity to rivals, local market features and demographics. 

\cite{reilly1998petrol} focus on the work of \cite{bacon1991rockets} and their findings do not support those of Bacon who claims the upward price process to be slightly faster and period of adjustment more concentrated than in case of downward price movement. According to \cite{reilly1998petrol}, both price changes are fully passed through in the long-run. \cite{eckert2002retail} studies the Canadian data (Windsor, Ontario) and rejects tacit collusion as the explanation of asymmetry, and instead points out that retail price series show an asymmetric cycle, which is not present in the wholesale prices series. And finally, \cite{kaufmann2005causes} argue that asymmetry is implied by the efficient gasoline markets so that there is little justification for policy interventions. 

\section{Data and preliminary analysis}

We analyze weekly gasoline prices for Belgium, France, Germany, Italy, the Netherlands, the UK and the USA and their possible asymmetric transmission referred to as the ``rockets and feathers'' effect in the literature. For the European markets, we utilize the Brent crude oil as an exogenous production variable, and for the US market, we use the WTI crude oil. The oil prices are the average weekly spot prices and the gasoline prices are the average retail prices for the given country. The whole dataset was obtained from www.eia.gov. The analyzed period starts at 08.01.1996 and goes up to 19.5.2014, which gives us 959 weekly prices for each variable.

\begin{figure}[htbp]
\center
\begin{tabular}{cc}
\includegraphics[width=3.1in]{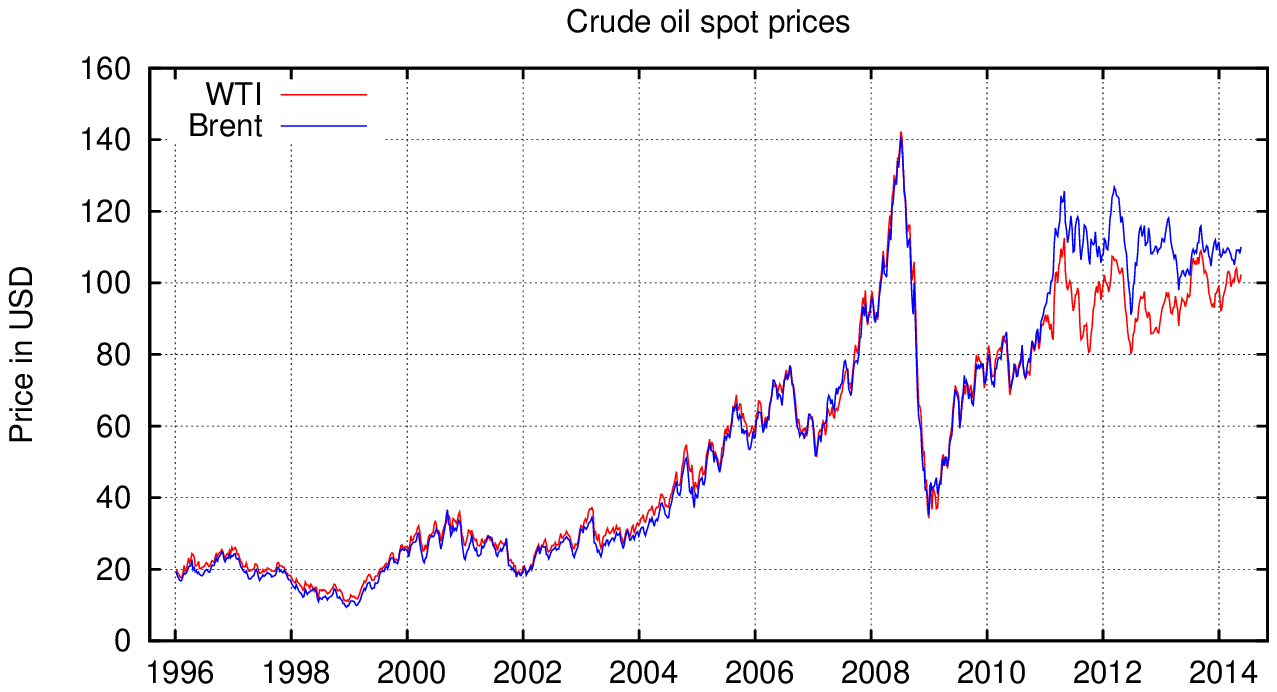}&\includegraphics[width=3.1in]{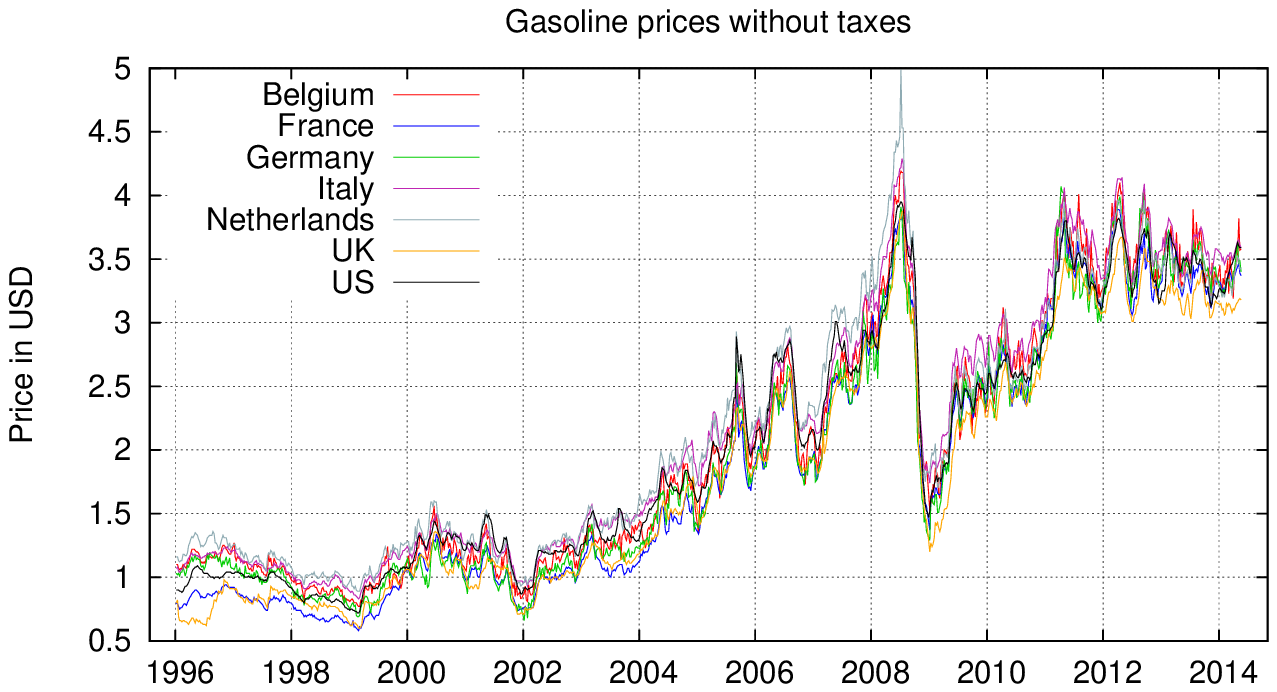}\\
\end{tabular}
\caption{\footnotesize\textbf{Crude oil and gasoline prices.} Crude oil prices (left panel) for Brent and WTI are reported in the US dollars per barrel. Retail gasoline prices (right panel) corrected for taxes are reported in the US dollars per gallon. All series have been obtained from www.eia.gov. \label{fig1}}
\end{figure}

Evolution of all the analyzed prices is illustrated in Fig. \ref{fig1}. The gasoline prices are reported without taxes and in the US dollars per gallon for better comparison. The crude oil prices are reported in the US dollars per barrel. We observe that the gasoline prices for all countries practically overlap for the whole analyzed period. The same thing can be said about the Brent and WTI crude oils up till 2011. However, from 2011 onwards, the WTI price remains below the Brent price due to changes in the US oil policies. Even though the initial divergence of the series is rather sharp, the prices have been converging during the last months.

Traditional ``rockets and feathers'' literature builds on the assumption that the gasoline and crude oil prices are cointegrated, i.e. they tend to a long-run equilibrium, in economic terms. As the crude oil price can be taken as an exogenous variable and all prices are reported in the US dollars, we can write the long-run equilibrium relationship as
\begin{equation}
\label{coint}
\log (G_{i,t})=\beta_0+\beta_1 \log (CO_{i,t}) + \varepsilon_{i,t}
\end{equation}
where $G_{i,t}$ is the gasoline price of country $i$ at time $t$, $CO_{i,t}$ is the crude oil price respective to country $i$ at time $t$. Due to the logarithmic specification of Eq. \ref{coint}, parameter $\beta_1$ can be interpreted as a long-term elasticity or a long-term passthrough. Error-correction term $\varepsilon_{i,t}$ is a deviation from the long-term equilibrium. If the prices return to their long-term equilibrium slower from above than from below, the situation is labelled as the ``rockets and feathers'' effect. Therefore, the analysis of behavior of the error-correction term separately above and below the equilibrium value becomes crucial.

As we have demonstrated in the preceding sections, the typical way how to approach such problem is to treat Eq. \ref{coint} as the cointegration relationship. If such relationship is found, the authors usually tackle the series using the (vector) error-correction model. Such procedure assumes that the original series $G$ and $CO$ are unit roots, i.e. integrated of order one, I(1), and the error-correction term is stationary and weakly dependent, i.e. integrated of order zero, I(0). As the cointegration relationship is usually built simply on the prices of gasoline and crude oil, the Engle-Granger two-step cointegration testing procedure is applied \citep{Engle1987}. The procedure stems in two steps. Firstly, the original series of Eq. \ref{coint} are tested for unit root using the Augmented Dickey-Fuller (ADF) test \citep{Dickey1979}. And secondly, if both the original series are found to contain unit root, the cointegration relationship in Eq. \ref{coint} is estimated using OLS and the residuals are tested for the unit root presence as well. If the unit root is rejected for the residuals, i.e. the estimated error-correction term, we say that series $G$ and $CO$ are cointegrated.

For our dataset, we find straightforward results for the original series, which are summarized in Tab. \ref{tab1} -- all the analyzed series contain unit root. In Tab. \ref{tab1}, we also apply the KPSS test \citep{Kwiatkowski1992}, which has a null hypothesis of I(0), i.e. weakly dependent stationarity. The latter test supports the finding of unit roots in all gasoline and crude oil prices. The series are thus eligible for possible cointegration relationships. Estimated long-run elasticities (also standardly referred to as passthrough or transmission), together with heteroskedasticity and autocorrelation consistent (HAC) standard errors, are summarized in Tab. \ref{tab2}. The transmissions vary between 0.6 and 0.8 and we thus do not observe a complete passthrough of the crude oil price and its changes into the gasoline prices for any of the analyzed markets. However, the more important findings are reported in the right part of the table.

\begin{figure}[htbp]
\center
\begin{tabular}{c}
\includegraphics[width=3.1in]{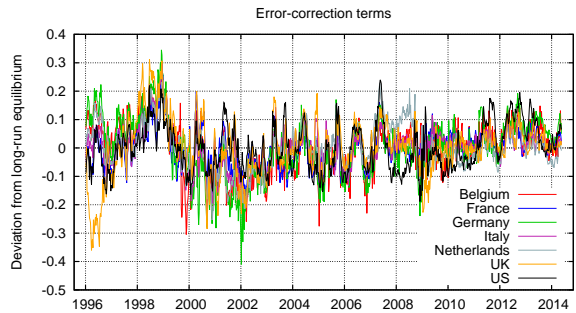}\\
\end{tabular}
\caption{\footnotesize\textbf{Error-correction terms.} Error-correction terms are obtained from the cointegration relationship between respective retail gasoline price and crude oil. Logarithmic relationship between the series is estimated based on Eq. \ref{coint}. \label{fig2}}
\end{figure}

There, we report the ADF and KPSS tests for the error-correction terms from all cointegration relationships, which are illustrated in Fig. \ref{fig2}. The results are again quite straightforward -- the error-correction terms are not unit root series but are also not stationary (or they are borderline stationary). This is further supported by the estimated $d$ parameters using the local Whittle \citep{Robinson1995} and the GPH \citep{Geweke1983} estimators. The estimates and standard errors suggest that most of the error-terms are borderline (non-)stationary with $0.5 \lesssim d < 1$. We can thus safely say that all of the analyzed pairs are cointegrated. However, we can also safely state that the error-correction terms are not I(0). Even though this does not play any significant part for the cointegration itself, it plays a crucial role in the usefulness of the (vector) error-correction models. Using the notation of Eq. \ref{coint}, we can write the error-correction model (ECM) as
\begin{equation}
\label{ECM}
\Delta \log (G_{i,t})=\gamma_0+\sum_{j=1}^{p}{\gamma_j\Delta \log (G_{i,t-j})}+\sum_{j=1}^{p}{\delta_j\Delta \log (CO_{i,t-j})}+\eta\hat{\varepsilon}_{i,t-1}+u_{i,t}.
\end{equation}
The regression is estimated using the ordinary least squares and parameter $\eta$ is negative for the cointegration relationship, i.e. the error-correction term reverts back to the mean values and the cointegrated pair does not diverge. The logarithmic differences of the gasoline and crude oil prices are I(0) automatically, i.e. from the fact that the prices are I(1). However, for a feasible estimation procedure, we also need stationary and weakly dependent error-correction term $\hat{\varepsilon}_t$. This is usually assumed from rejection of the null hypothesis of the ADF test, i.e. from rejection of unit root. However, rejection of I(1) does not automatically imply either of I(0), stationarity or weak dependence. Figures shown in Tab. \ref{tab2} clearly show that the error-correction term does not meet the necessary criteria for ECM to be correctly estimated. There are various reasons why the estimation procedure in Eq. \ref{ECM} does not work when the error-correction term is not I(0). We now shortly focus on the most obvious one.

Assume that Eq. \ref{ECM} holds and also assume that the error-correction term $\hat{\varepsilon}_{i,t-1}$ is integrated of order higher than zero, i.e. it is a long-term memory process. From the definition of the standard cointegration relationship, we know that both $\log(G_{i,t})$ and $\log(CO_{i,t})$ are integrated of order one, i.e. I(1). Their first differences are thus automatically I(0). Turning now back to Eq. \ref{ECM}, we have an I(0) process (left hand side of the equation) being a sum of three I(0) processes (gasoline, crude oil and an error term $u_{i,t}$) and one process integrated of order higher than zero. This is a contradiction as the sum of integrated processes is asymptotically integrated of the same order as the highest order among the separate processes \citep{Samorodnitsky2006,Kristoufek2013}. The estimation is thus inconsistent.

Even though we do not replicate the time series analyzed in other studies using ECM and asymmetric ECM, we can quite confidently speculate that the statistical and dynamic properties of the gasoline and crude oil series do not differ much from the ones we report and it is very likely that the same problem exists even for other studies. Application of ECM (or the asymmetric ECM which is popular in the ``rockets and feathers'' literature) thus yields unreliable results. Any study dealing with the asymmetric passthrough from crude oil to gasoline prices using the cointegration framework should take this issue into consideration. In the next section, we introduce three tests which build on the cointegration methodology and possible asymmetry of the error-correction term. The tests are constructed using the characteristics of the mean-reverting time series and they do not need the analyzed series to be either I(0) or stationary or weakly dependent. 

\section{Methodology}

Cointegration framework is a natural environment for analyzing the price transmission from crude oil to retail gasoline. The ``rockets and feathers'' dynamics of the relationship can be simply understood as the fact that it takes prices a longer time before they converge back to their equilibrium level if gasoline is overpriced (with respect to the cointegration long-term equilibrium) than if it is underpriced. In the previous section, we have shown that the error-correction term, which represents such deviation from the equilibrium state, is fractionally integrated of order less than one which implies that the term is mean-reverting and the gasoline price thus returns to its equilibrium level. We can use the mean reversion approach in the ``rockets and feathers'' framework by saying that if the effect is existent on the specific market, then the positive part of the error-correction term will revert to its mean more slowly than the negative part. In this section, we introduce three new tests based on this idea.

\subsection{Median test}

The simplest idea of mean-reverting processes asymmetric around the equilibrium value is that one of the two parts of the series stays more on one side of the mean. Even though the mean value of the error-correction term is given to be zero, its median would standardly not be. Taking the median value of the error-correction term $\hat{\varepsilon}_t$ based on Eq. \ref{coint} as a measure of asymmetry which is robust to extreme movements, we construct a simple testing statistic $M=median(\{\hat{\varepsilon}_t\}_{t=1}^T)$. For the symmetric error-correction term dynamics, the median is close to zero but for the ``rockets and feathers'' effect, the series should be skewed to positive values and the median will be positive. To be able to make any claims statistically valid, we construct the critical values based on surrogate Fourier randomized series of $\hat{\varepsilon}_t$. As the Fourier randomized series keep the spectral characteristics of the series, they serve as a good approximation of the null hypothesis. The testing statistic $M$ is calculated for each of the 10,000 randomized series to form $M^{\ast}$. For a significance level $\alpha$, we find the critical value $M^{\ast}_{\alpha}$ of the one-sided alternative (the ``rockets and feathers'' effect) as the $1-\alpha$ quantile of $M^{\ast}$ and we reject the null hypothesis if $M>M_{\alpha}^{\ast}$.

\subsection{Wave test}

Mean-reverting persistent time series are characteristic by wandering quite far away from the mean value and for long time periods. Labeling the values above mean as $+$ and the values below mean as $-$, we can obtain a series such as $+++----++-$ which consists of four runs -- two positive ones with lengths of three and two, and two negative ones with lengths of four and one. Let's say that we have a set of positive runs with given lengths $W^+$ and a set of negative runs with given lengths $W^-$. In the example, we have $W^+\in \{2,3\}$ and $W^-\in \{1,4\}$.

Let's return to the case of error-correction term and its possible asymmetry around mean. In the case of symmetry, series both above and below mean have the same mean-reversion rate so that the length of runs should be on average the same. In the case of the ``rockets and feathers'' dynamics, the error-correction term should stay longer above its mean value before it returns to its equilibrium level than if it's below its mean value. Utilizing this characteristic, we propose a new test based on a difference between the average length of runs above and below the mean value. As the wandering away from the mean value is rather persistent for this specific case, we rather refer to these persistent runs as waves. This way, we also distinguish between standard runs tests, which are used to test no serial correlation of the series whereas the waves test examines potential asymmetry in the dynamics around the mean value.

The wave testing statistic $W$ is defined as
\begin{equation}
\label{Wave}
W=\overline{W^+}-\overline{W^-}
\end{equation}
where $W^+$ is an average length of the positive runs in the error-correction term $\hat{\varepsilon}_t$ and $W^-$ is an average length of the negative runs. For the symmetric error-correction term, the expected value of the $W$ statistic is zero whereas for the prevailing positive runs, i.e. the slower mean-reversion of the values above the equilibrium state which corresponds to the ``rockets and feathers'' effect, the statistic is positive. The null hypothesis of a symmetric error-correction term against the alternative of the ``rockets and feathers'' effect is again based on Fourier randomization of the error-correction term $\hat{\varepsilon}_t$ obtained from Eq. \ref{coint}. Specifically, we take the Fourier randomized series of $\hat{\varepsilon}_t$, which share spectral properties of the original series and it is thus symmetric, and calculate $W^{\ast}$ based on Eq. \ref{Wave}. Such procedure is repeated 10,000 times and the distribution of the $W$ statistic under the null hypothesis is obtained. For a selected significance level $\alpha$, we find a one-sided critical value $W^{\ast}_{\alpha}$ as the $1-\alpha$ quantile of the $W^{\ast}$ distribution. The null hypothesis is rejected, i.e. the ``rockets and feathers'' effect is found, if $W>W^{\ast}_{\alpha}$.

\subsection{Rescaled range ratio test}

In the previous section, we show that the error-correction terms for all analyzed series are non-stationary or borderline stationary. Specifically, the fractional differencing parameter $d$ is very far from the assumed $d=0$ assumed for standard error-correction models. Even if $d \gtrsim 0.5$ for all series, which disqualifies the use of standard error-correction models, the notion of fractional integration and long-term memory still provides ways to test for asymmetry in the error-correction term dynamics around its mean. The higher the $d$ parameter is, the more persistent the underlying series is and thus the more it wanders away from its long-term mean value. Therefore, we assume that the level of persistence is the same for both parts (positive and negative) of the symmetric error-correction term. And for the ``rockets and feathers'' asymmetry, we would observe that the positive part of the error-correction term is more persistent than the negative part.

However, it turns out that testing for difference in the fractional integration parameters $d$ of part of one series is much more troublesome than testing the difference between two series. This is mainly due to the nature of the error-correction term $\hat{\varepsilon}_t$ separation into two series -- the positive and the negative ones. The positive part takes the same values of the original series if these are positive and zero otherwise, and symmetrically for the negative part. Each of these series thus has long periods when being equal to zero. This levies strong autocorrelation structure into the series so that we cannot simply estimate the $d$ parameters of the separate series and compare these. We cannot even use the two-sample test of \cite{Lavancier2010} which is specifically constructed for testing equality of $d$ of two series. To overcome these issues, we introduce a new test.

Motivated by the test of \cite{Lavancier2010} which is based on the univariate rescaled variance test of \cite{Giraitis2003}, we propose a parallel test based on the rescaled range test originally utilized by \cite{hurst1951} and later studied and popularized by Beno\^{\i}t Mandelbrot \citep{Mandelbrot1968,Mandelbrot1971,Mandelbrot1972}. Similarly to the original method, we construct a range of the series' profile, i.e. a difference between maximum and minimum of the cumulative deviations from mean. However, our series have specific properties and the aim of the test is different so that we need to alter the original methodology.

We construct ranges for each part of the error-correction term $\hat{\varepsilon}_t$ and we label them as $R^+$ and $R^-$ for the positive part and the negative part, respectively. Formally, this is expressed as

\begin{equation}
\label{Rplus}
R^+=\max\left(\sum_{t=1}^T{\hat{\varepsilon_t}\mathbb{I}_{\hat{\varepsilon_t}\ge 0}}\right)-\min\left(\sum_{t=1}^T{\hat{\varepsilon_t}\mathbb{I}_{\hat{\varepsilon_t}\ge 0}}\right)=\sum_{t=1}^T{\hat{\varepsilon_t}\mathbb{I}_{\hat{\varepsilon_t}\ge 0}}
\end{equation}

\begin{equation}
\label{Rminus}
R^-=\max\left(\sum_{t=1}^T{\hat{\varepsilon_t}\mathbb{I}_{\hat{\varepsilon_t}< 0}}\right)-\min\left(\sum_{t=1}^T{\hat{\varepsilon_t}\mathbb{I}_{\hat{\varepsilon_t}< 0}}\right)=-\sum_{t=1}^T{\hat{\varepsilon_t}\mathbb{I}_{\hat{\varepsilon_t}<0}}
\end{equation}
where $\mathbb{I}_{\bullet}$ is an indicator function equal to 1 if the condition in $\bullet$ is met and 0 otherwise. To take into consideration the fact that the scale of each part differs, we rescale each range using its variance. However, as the series are constructed as the negative and the positive part of the error-correction term, standard variance would introduce bias through its estimated mean value. To control for this specific, we utilize semi-variances of the series rather than variances. If the error-correction term varies symmetrically around its mean value, the rescaled ranges of each part should be the same (asymptotically). In the case of the ``rockets and feathers'' asymmetry, the rescaled range of the positive part should dominate the other one. This leads us to construct the testing statistic, which we label as the rescaled range ratio (RRR) statistic, as

\begin{equation}
\label{RRR}
RRR=\frac{R^{+}}{R^{-}}\times \frac{\sum_{t\in \{t: \hat{\varepsilon}_t<0\}}{\hat{\varepsilon}_t^2}}{\sum_{t\in \{t: \hat{\varepsilon}_t\ge0\}}{\hat{\varepsilon}_t^2}}.
\end{equation}
In the same way as for the previous two tests, the rejection of the null hypothesis of the error-correction term symmetry is based on the Fourier randomized surrogate series. For a given significance level $\alpha$, we again reject the null hypothesis in favor of the alternative of the ``rockets and feathers'' if testing statistic for the original series $RRR$ exceeds the critical level $RRR_{\alpha}^{\ast}$ based on 10,000 Fourier randomized replications.

\section{Application and discussion}

In the Data and preliminary analysis section, we have shown that for all the studied gasoline markets (Belgium, France, Germany, Italy, the Netherlands, the UK, and the US), the relationship with the given crude oil (either Brent or WTI) is identified as a cointegration one, i.e. the gasoline and crude oil prices tend to an equilibrium value. The price transmission from crude oil to gasoline varies approximately between 0.6 and 0.8 so that it is quite strong yet still imperfect. Deviations from the long-term equilibrium gasoline prices, which are represented by the error-correction term, have been shown to deviate significantly from an I(0) process. Such dynamics can be also observed by a naked eye in Fig. \ref{fig2}. The term is thus not weakly dependent and mostly on the verge of (non-)stationarity. Further, we have shown that such error-correction term makes standard ECM models invalid. To be able to use the cointegration framework for distinguishing between symmetric and asymmetric dynamics of the error-correction term, we have introduced three new tests in the previous section. Results of the tests now follow.

Tab. \ref{tab3} summarizes all the results and it includes the testing statistics and $p$-values for the null hypothesis of symmetric adjustment of the error-correction term coming from Eq. \ref{coint} against the one-sided alternative hypothesis of the ``rockets and feathers'' asymmetry, i.e. the case when the error-correction term reverts to the equilibrium more slowly when its above equilibrium compared to the situation when its below the equilibrium value. The $p$-values are based on 10,000 Fourier randomized series as the null hypothesis holds for these randomized series.

The median test provides a straightforward evidence that we can find asymmetry only for Belgium and this is on the 5\% level. Other median testing statistics do not show significant results. Interestingly, we report negative values for the Dutch and the US markets which would suggest a reversed asymmetry. However, we are primarily interested in reinvestigating the standard ``rockets and feathers'' effect and we leave possibility of the inverse relationship for further research.

For the wave test, we again find the market of Belgium to behave according to the ``rockets and feathers'' dynamics. Moreover, we also find Italy to be significantly asymmetric at 10\% level. Other markets are statistically insignificant. Again, the wave test reports negative values for the Netherlands and the US.

And the rescaled range ratio test further supports the Belgian market as the asymmetric one. No other market gives significant results. Again, we have the Dutch and the US markets signifying the inverse effect with the ratio below one.

Summarizing, we find that apart from the gasoline market of Belgium, there is no statistically significant evidence of the ``rockets and feathers'' effect. This could be attributed to the size of the country as argued by \cite{duffy1996retail}, \cite{verlinda2008rockets} and \cite{deltas2008retail}. So far the role of the state in the ``rockets and feathers'' issue has been limited to antitrust investigations e.g. (significant) market power abuse or tacit collusion. According to our results, the states' approach seems to be correct as the ``rockets and feathers'' dynamics has not been proven on national level (with the exception of Belgium). However, it may apply in specific local markets. In addition, the results signify that some of the markets, specifically the Dutch and the US ones, might experience the inverse effect. We thus do not find the problematic asymmetry, i.e. the one which is hardly received by customers, on most of the markets. Nevertheless, the research on the topic is of course not complete.

Firstly, we have found no asymmetry at national level. However, more localized study could report qualitatively different results. Secondly, the set of tests we have newly introduced in this article does not cover all possibilities. There are some other approaches that could be added such as fractionally integrated ECM or fractional cointegration framework in general. And thirdly, we do not investigate various stages of the price transmission. The article thus primarily serves as a starting point for treating the asymmetric equilibrium adjustment of the error-correction term in a different, statistically and econometrically convenient, way. Finally, it has to be noted that the developed tests are not restricted to the relationship between retail gasoline, crude oil and related variables but they can serve to test the asymmetry in any economic and financial application which considers asymmetry in the cointegration framework.

\section{Conclusion}

We have analyzed the possible ``rockets and feathers'' dynamics between the retail gasoline and crude oil prices. Focusing on the national prices of selected countries, we provide a step-by-step treatment in the cointegration framework. The standardly applied error-correction model methodology is discussed in detail and we show that it is not convenient for the analysis of asymmetry in the price transmission in the given system due to long-term memory aspects of the equilibrium adjustments represented by the error-correction term. We show that the gasoline prices return to their equilibrium levels much more slowly than assumed by the ECM approach which makes the estimation inconsistent and the results are thus unreliable. To deal with such issue, and to still remain in the cointegration environment, we introduce three new tests for asymmetry in the error-correction term -- median test, wave test and rescaled range ratio test.

On the dataset of seven national gasoline price series, we show that apart from Belgium, we find no statistically significant signs of the ``rockets and feathers'' effect. However, this does not necessarily discard the previous results showing asymmetry as we limit ourselves on the national data only. The results might indeed differ for more local price series.

Importantly, the proposed framework is not limited only to the gasoline-oil relationship but it can be utilized for any economic and financial series which are considered in the equilibrium cointegrated relationship and the adjustment rate might be asymmetric. The article can thus serve as a reference for future research in this area.

\section*{Acknowledgements}

Support from the Czech Science Foundation under projects No. P402/11/0948 and No. 14-11402P is gratefully acknowledged.

\section*{Supplementary files}
Scripts (written for R 3.0.0) for the median test, the wave test and rescaled range ratio test are appended to the manuscript. Data sets used in the analysis are publicly available at http://www.eia.gov. The specific analyzed series are appended to the manuscript as well. Several missing observations in the gasoline series have been filled in using linear projection from the nearest neighbors.

\section*{References}

\bibliographystyle{chicago}
\bibliography{Rockets_EE}

\newpage

\begin{landscape}

\begin{table}[htbp]
\centering
\caption{Summary of the ``rockets and feathers'' literature}
\label{tabLR}
\scriptsize
\begin{tabular}{ccccc}
\toprule \toprule
Reference&Period&Country&Model/method&Results\\
\midrule
\cite{al2007retail} & 1998--2004 & USA&TAR, M-TAR, VECM & Asymmetry\\
\cite{bachmeier2003new} & 1985--1998 & USA & ECM (asymmetric)& Symmetry\\
\cite{bacon1991rockets} & 1982--1989 & UK  & Quadratic quantity adjustment function& Asymmetry \\
\cite{balke1998crude} & 1987--1996 & USA &  ECM (asymmetric) & Asymmetry\\
\cite{balmaceda2008asymmetric} & 2001--2004 & Santiago, Chile  & ECM & Asymmetry\\
\cite{bettendorf2003price} & 1996--2001 & Netherlands &  ECM (asymmetric) & Neutral \\
\cite{borenstein2002sticky}&1985--1995 & USA &  LAM, PAM and VAR& Asymmetry\\
\cite{borenstein1997gasoline} & 1986--1992 & USA  & ECM & Asymmetry\\
\cite{chen2005threshold} & 1991--2003 & USA & ECM (threshold) & Asymmetry\\
\cite{deltas2008retail}&1988--2002&USA (separate states) &  ECM (various) & Asymmetry\\
\cite{douglas2010gasoline}&1990--2008 & USA &  ECM & Depends on outliers\\
\cite{duffy1996retail}&1989--1993 & Salt Lake City, USA & Markup model with first differences & Asymmetry\\
\cite{eckert2002retail}&1989--1994 & Windsor, Ontario, Canada  & ECM (reduced) & Asymmetry\\
\cite{galeotti2002rockets}&1985--2000 & International (DE, ES, FR, IT, UK) &  ECM (dynamic) & Asymmetry\\
\cite{godby2000testing}&1990--1996 & Canada (13 cities) & TAR within EC framework&Symmetry\\
\cite{grasso2007asymmetric}&1985--2003& International (DE, ES, FR, IT, UK) & ECM (asymmetric, threshold) & Asymmetry\\
\cite{honarvar2009asymmetry}&1981--2007 & USA & ECM (crouching) & Asymmetry\\
\cite{johnson2002search}&1996--1998 & USA (15 cities)& ECM & Asymmetry\\
\cite{karrenbrock1991behavior}&1983--1990 & USA & Markup model with first differences & Symmetry\\
\cite{kaufmann2005causes}&1986--2002 & USA & ECM (restricted and unrestricted) & Asymmetry\\
\cite{lewis2011asymmetric}& 2000--2001 & San Diego, CA, USA &Consumer search model (with EC term) & Asymmetry\\
\cite{liu2010there}& 2004--2009 & New Zealand & ECM (asymmetric) & Asymmetry \\
\cite{nagy1998asymmetry}& 1980--1996 & UK and USA & ECM (dynamic) & Asymmetry\\
\cite{oladunjoye2008market}&1987--2004 & USA & ECM (asymmetric) & Symmetry\\
\cite{panagiotidis2007oil}&1996--2003 & UK & VECM & Symmetry\\
\cite{radchenko2005oil}& 1993--2003 & USA & ECM, VAR and PAM& Asymmetry\\
\cite{reilly1998petrol}&1982--1995 & UK & ECM (unrestricted dynamic) & Asymmetry \\
\cite{tappata2009rockets} & Theoretical & General & Consumer search model & Asymmetry\\
\cite{verlinda2008rockets}& 2002--2003 & USA & ECM & Asymmetry\\
\bottomrule \bottomrule
\end{tabular}
\textit{Abbreviations: ECM (error-correction model), M-TAR (momentum threshold autoregressive model), PAM (partial adjustment model), LAM (lagged adjustment model), TAR (threshold autoregressive model), VAR (vector autoregession), VECM (vector error-correction model)}
\end{table}

\end{landscape}

\begin{table}[htbp]
\centering
\caption{Unit-root and stationarity testing}
\label{tab1}
\footnotesize
\begin{tabular}{c|cccc}
\toprule \toprule
Country&ADF&$p$-value&KPSS&$p$-value\\
\midrule \midrule
Belgium&-1.1166&$>0.1$&10.9231&$<0.01$\\
France&-1.1415&$>0.1$&11.1559&$<0.01$\\
Germany&-1.1823&$>0.1$&10.8568&$<0.01$\\
Italy&-0.9969&$>0.1$&11.1836&$<0.01$\\
Netherlands&-1.2177&$>0.1$&10.6809&$<0.01$\\
UK&-1.6644&$>0.1$&10.9663&$<0.01$\\
US&-0.8690&$>0.1$&11.0235&$<0.01$\\
\midrule
Brent &-0.9535&$>0.1$&11.0037&$<0.01$\\
WTI&-1.1199&$>0.1$&10.8473&$<0.01$\\
\bottomrule \bottomrule
\end{tabular}
\end{table}

\begin{table}[htbp]
\centering
\caption{Cointegration \& error-correction term testing}
\label{tab2}
\footnotesize
\begin{tabular}{c|cc||cc|cc|cc}
\toprule \toprule
Country&Transmission&SE&ADF&$p$-value&KPSS&$p$-value&LWE&GPH\\
\midrule \midrule
Belgium&0.6804&0.0123&-3.2763&0.0160&1.2094&$<0.01$&0.6574 [0.0645]&0.6857 [0.0924]\\
France&0.7842&0.0096&-5.1817&$<0.01$&0.7797&$<0.01$&0.5201 [0.0645]&0.6012 [0.0982]\\
Germany&0.7005&0.0146&-3.7932&$<0.01$&1.2647&$<0.01$&0.7139 [0.0645]&0.8044 [0.1117]\\
Italy&0.6690&0.0104&-2.8726&0.0486&0.8037&$<0.01$&0.6138 [0.0645]&0.6709 [0.1102]\\
Netherlands&0.6329&0.0114&-3.9637&$<0.01$&0.5147&0.0420&0.7105 [0.0645]&0.7000 [0.1024]\\
UK&0.7478&0.0146&-4.0413&$<0.01$&0.3121&$>0.1$&0.6036 [0.0645]&0.5217 [0.0936]\\
US&0.7560&0.0106&-4.4462&$<0.01$&0.6123&0.0290&0.4630 [0.0645]&0.4995 [0.1275]\\
\bottomrule \bottomrule
\end{tabular}
\end{table}

\begin{table}[htbp]
\centering
\caption{Asymmetry in error-correction term testing}
\label{tab3}
\footnotesize
\begin{tabular}{c|cccccc}
\toprule \toprule
Country&Median test&$p$-value&Wave test&$p$-value&RRR test&$p$-value\\
\midrule \midrule
Belgium&0.0102&0.0234&1.1351&0.0184&1.2454&0.0581\\
France&0.0012&$>0.1$&0.2308&$>0.1$&0.8675&$>0.1$\\
Germany&0.0024&$>0.1$&0.2511&$>0.1$&1.0822&$>0.1$\\
Italy&0.0059&$>0.1$&1.2083&0.0743&1.0360&$>0.1$\\
Netherlands&-0.0042&$>0.1$&-0.7736&$>0.1$&0.9311&$>0.1$\\
UK&0.0026&$>0.1$&0.2958&$>0.1$&1.1218&$>0.1$\\
US&-0.0069&$>0.1$&-1.5098&$>0.1$&0.7763&$>0.1$\\
\bottomrule \bottomrule
\end{tabular}
\end{table}

\end{document}